\documentclass[amsfonts,tightenlines,aps,12pt,floatfix,nofootinbib]{revtex4}
\usepackage{bm}
\usepackage{epsfig}
\usepackage{float}
\usepackage{color}
\def\ltap{\raisebox{-.4ex}{\rlap{$\sim$}} \raisebox{.4ex}{$<$}}   % < or ~
\def\gtap{\raisebox{-.4ex}{\rlap{$\sim$}} \raisebox{.4ex}{$>$}}   % > or ~
\begin{document}
% Use the \preprint command to place your local institutional report
% number in the upper righthand corner of the title page in preprint mode.
% Multiple \preprint commands are allowed.
% Use the 'preprintnumbers' class option to override journal defaults
% to display numbers if necessary
%\preprint{}
\thispagestyle{empty}

%Title of paper
\title{A New Strategy for the Lattice Evaluation of the Leading
Order Hadronic Contribution to \begin{boldmath}$(g-2)_\mu$\end{boldmath}}

\author{Maarten Golterman}
\email[]{maarten@sfsu.edu}
\affiliation{Department of Physics and Astronomy,
San Francisco State University, San Francisco, CA 94132, USA}
\author{Kim Maltman}
\email[]{kmaltman@yorku.ca}
\affiliation{Department of Mathematics and Statistics, York University,
4700 Keele St., Toronto, ON CANADA M3J 1P3}
\altaffiliation{CSSM, Univ. of Adelaide, Adelaide, SA 5005 AUSTRALIA}
\author{Santiago Peris}
\email[]{peris@ifae.es}
\affiliation{Department of Physics, Universitat Aut\`onoma de Barcelona,
\\ E-08193 Bellaterra, Barcelona, Spain}

\begin{abstract}
A reliable evaluation of the integral giving the
hadronic vacuum polarization contribution to the muon anomalous magnetic
moment should be possible using a simple trapezoid-rule integration of lattice
data for the subtracted electromagnetic current polarization function in
the Euclidean momentum interval $Q^2>Q^2_{min}$, coupled with an
$N$-parameter Pad\'e or other representation of the
polarization in the interval $0<Q^2<Q^2_{min}$, for  sufficiently high
$Q^2_{min}$ and sufficiently large $N$.
Using a physically motivated model for the $I=1$ polarization, and
the covariance matrix from a recent lattice simulation to generate
associated fake ``lattice data,'' we show that systematic errors associated
with the choices of $Q^2_{min}$ and $N$ can be reduced to well below the
$1\%$ level for $Q^2_{min}$ as low as $0.1$~GeV$^2$ and rather small
$N$. For such low $Q^2_{min}$, both
an NNLO chiral representation with one additional NNNLO term and a
low-order polynomial expansion employing a conformally transformed variable
also provide representations sufficiently accurate to reach this
precision for the low-$Q^2$ contribution.
Combined with standard techniques for reducing other
sources of error on the lattice determination, this hybrid
strategy thus looks to provide
a promising approach to reaching the goal of a sub-percent precision
determination of the hadronic vacuum polarization contribution
to the muon anomalous magnetic moment on the lattice.

\end{abstract}

% insert suggested PACS numbers in braces on next line
%\pacs{12.38.Gc, 13.40.Em, 11.55.Fv}
% insert suggested keywords - APS authors don't need to do this
%\keywords{}

\maketitle

\section{\label{intro}Introduction}
The discrepancy of about $3.5\sigma$ between the measured value~\cite{bnlgminus2}
and Standard Model prediction~\cite{SMgminus2ref} for the anomalous
magnetic moment of the muon, $a_\mu = (g_\mu -2)/2$, has attracted
considerable attention. After the purely QED contributions, which are now
known to five loops~\cite{kinoshita5loopqed}, the next most important term
in the Standard Model prediction is the leading order (LO) hadronic vacuum polarization
(HVP) contribution, $a_\mu^{\rm LO,HVP}$. The error on the dispersive evaluation
of this quantity, obtained from the errors on the input
$e^+ e^-\rightarrow hadrons$ cross-sections, is currently the largest of
the contributions to the error on the Standard Model prediction~\cite{SMgminus2ref}.
The dispersive approach is, moreover, complicated by discrepancies between
the determinations by different experiments of the cross-sections for the
most important exclusive channel,
$e^+e^-\rightarrow\pi^-\pi^+$~\cite{CMD2pipi07,SNDpipi06,BaBarpipi12,KLOEpipi12}.{\footnote{A
useful overview of the experimental situation is given
in Figs.~48 and 50 of Ref.~\cite{BaBarpipi12}.}}

The existence of this discrepancy, and the role played by the error on the
LO HVP contribution, have led to an increased interest in providing an
independent  determination of
$a_\mu^{\rm LO,HVP}$ from the lattice
\cite{TB12,TB03,AB07,FJPR11,BDKZ11,DJJW12,abgp12,DPT12,FHHJPR13,FJMW13,ABGP13,BFHJPR13,gmp13,HHJWDJ13,hpqcd14}.
Such a determination is made possible by the representation of
$a_\mu^{\rm LO,HVP}$ as a weighted integral of the subtracted polarization,
$\hat{\Pi}(Q^2)$, over Euclidean momentum-squared $Q^2$~\cite{TB03,ER}.
Explicitly,
\begin{eqnarray}
\label{amu}
a_\mu^{\rm LO,HVP}&=&-4\alpha^2\int_0^\infty dQ^2\,f(Q^2)\,
{\hat{\Pi}}(Q^2)\, , \end{eqnarray}
where, with $m_\mu$ the muon mass,
\begin{eqnarray}
f(Q^2)&=&m_\mu^2 Q^2 Z^3(Q^2)\,\frac{1-Q^2 Z(Q^2)}{1+m_\mu^2 Q^2
  Z^2(Q^2)}\ ,\nonumber\\
Z(Q^2)&=&\left(\sqrt{(Q^2)^2+4m_\mu^2 Q^2}-Q^2\right)/
(2m_\mu^2 Q^2)\ ,
\end{eqnarray}
and $\hat{\Pi}(Q^2)\equiv\Pi (Q^2)-\Pi (0)$, with $\Pi (Q^2)$
the unsubtracted polarization, defined from the hadronic electromagnetic
current-current two-point function, $\Pi_{\mu\nu}(Q)$, via
\begin{equation}
\label{Pem}
\Pi_{\mu\nu}(Q)=\left(Q^2\delta_{\mu\nu}-Q_\mu Q_\nu\right)\Pi (Q^2)\ .
\label{polndefn}\end{equation}
The vacuum polarization
$\Pi_{\mu\nu}(Q)$ can be computed, and hence $\Pi (Q^2)$ determined
for non-zero $Q$, for those quantized Euclidean $Q$ accessible on a
given finite-volume lattice. Were $\Pi (Q^2)$ to be determined on
a sufficiently finely spaced $Q^2$ grid, especially in the region of
the peak of the integrand,  $a_\mu^{\rm LO,HVP}$ could be determined from lattice data
by direct numerical integration.

Two facts complicate such a determination.
First, since the kinematic tensor on the RHS of Eq.~(\ref{polndefn}),
and hence the entire two-point function signal, vanishes as $Q^2\rightarrow
0$, the errors on the direct determination of $\Pi (Q^2)$ become very large
in the crucial low-$Q^2$ region. Second, for the lattice
volumes employed in current simulations, only a limited number of points
is available in the low-$Q^2$ region, at least for conventional
simulations with periodic boundary conditions.   With the peak of the
integrand centered
around $Q^2\sim m_\mu^2/4\approx 0.0028$~GeV$^2$, one would
need lattices with a linear size of about 20~fm to obtain lattice data near
the peak.

The rather coarse coverage and sizable errors at very low $Q^2$ make it necessary to fit the lattice
data for $\Pi (Q^2)$ to some functional form, at least in the low-$Q^2$ region.
Existing lattice determinations have typically attempted to fit the form of
$\Pi (Q^2)$ over a sizable range of $Q^2$, a strategy partly
predicated on the fact that the errors on the lattice determination
are much smaller
at larger $Q^2$, and hence more capable of constraining the parameters of
a given fit form. The necessity of effectively extrapolating high-$Q^2$,
high-acccuracy data to the low-$Q^2$ region most relevant to
$a_\mu^{\rm LO,HVP}$
creates a potential systematic error difficult to quantify using lattice
data alone.

In Ref.~\cite{gmp13}, this issue was investigated using a physical model
for the subtracted $I=1$ polarization, $\hat{\Pi}^{I=1}(Q^2)$. The model
was constructed using the dispersive representation of $\hat{\Pi}^{I=1}(Q^2)$,
with experimental hadronic $\tau$ decay data used to
fix the relevant input spectral function. The study showed that (1)
$\hat{\Pi}^{I=1} (Q^2)$ has a significantly stronger curvature at low
$Q^2$ than at high $Q^2$ and (2), as a result, the extrapolation to
low $Q^2$ produced by typical lattice fits, being more
strongly controlled by the numerous small-error large-$Q^2$ data
points, is systematically biased towards producing insufficient
curvature in the low-$Q^2$ region either not covered by the data,
or covered only by data with much larger errors. Resolving this problem
requires an improved focus on contributions from the low-$Q^2$
region and a reduction in the impact of the large-$Q^2$ region on
the low-$Q^2$ behavior of the fit functions and/or procedures employed.

In this paper we propose a hybrid strategy to accomplish these goals.
The features of this strategy are predicated on a study of the $I=1$
contribution to $a_\mu^{\rm LO,HVP}$ corresponding to the model for the
$I=1$ polarization function, $\hat{\Pi}^{I=1}(Q^2)$, introduced in Ref.~\cite{gmp13}.  The results of this
study lead us to advocate a combination of direct numerical integration of
the lattice data in the region above $Q^2_{min}\sim 0.1$~GeV$^2$, and the
use of Pad\'e or other representations in the low-$Q^2$ ($0<Q^2\le Q^2_{min}$)
region. We will consider two non-Pad\'e alternatives for representing
$\hat{\Pi}$ at low $Q^2$, that provided by chiral perturbation theory
(ChPT) and that provided by a polynomial expansion in a conformal transformation of the variable $Q^2$ improving the convergence properties of the expansion.

The organization of the paper is as follows.
In Sec.~\ref{sec2} we briefly review the construction of the model,
and use the resulting $\hat{\Pi}^{I=1}(Q^2)$ to quantify expectations
about both the behavior of the integrand for
$\hat{a}_\mu^{\rm LO,HVP}\equiv \left[ a_\mu^{\rm LO,HVP}\right]^{I=1}$
and the accumulation of contributions to this quantity as
a function of the upper limit of integration in the analogue of
Eq.~(\ref{amu}). We also show, with fake data generated from the
model using the covariances and $Q^2$ values of a typical
lattice simulation with periodic boundary conditions, that
the contribution to $\hat{a}_\mu^{\rm LO,HVP}$ from $Q^2$ above $Q^2_{min}$
can be evaluated with an error well below $1\%$ of the full contribution
by direct trapezoid-rule numerical integration for $Q^2_{min}$ down to at
least as low as $Q^2_{min}=0.1$~GeV$^2$. The values of $Q^2$ covered by
state-of-the-art lattice data are too few, and the statistical
errors too large, to allow $Q^2_{min}$ to be lowered much
beyond this at present.
Such a low $Q^2_{min}$, however, implies that the use of fit forms
to represent the polarization function below $Q^2_{min}$ can be restricted
to the  region $Q^2\,\ltap\, 0.1-0.2$~GeV$^2$, where
the behavior of $\hat{\Pi}^{I=1}(Q^2)$ is expected to be much easier
to parametrize in a simple and reliable manner.
We then show, in Sec.~\ref{sec3}, that this expectation is borne out
in practice. Explicitly, we demonstrate that, in the region up to
about $0.1-0.2$~GeV$^2$, good enough data will
allow $\hat{\Pi}^{I=1}(Q^2)$ to be represented with an accuracy sufficient
to reduce the systematic error on the low-$Q^2$ contribution to
$\hat{a}_\mu^{LO,HVP}$ to well below the $1\%$ level.   The three
functional forms we investigate are
low-order Pad\'e's, a polynomial representation in a conformally
mapped variable, and
a next-to-next-to-leading-order (NNLO)
ChPT form supplemented by an analytic NNNLO term.
The Pad\'e's we will consider are of two types: those constrained
explicitly to reproduce the first few derivatives at $Q^2=0$
\cite{hpqcd14}, and those obtained by fitting to data in the low-$Q^2$
region \cite{abgp12}. We will be limited to investigating the
systematics of these low-$Q^2$ representations.
The lattice $Q^2$ values and covariance matrix employed for fake-data
studies in Ref.~\cite{gmp13} do not allow for a meaningful
extension of this exploration to
include also the statistical component of the uncertainty.
We expect, however, that new lattice data, employing
twisted boundary conditions to provide a denser set
of $Q^2$ values on the
lattice \cite{DJJW12,ABGP13,HHJWDJ13},
as well as improved statistics \cite{bis12,amaref}, will make a more
complete investigation possible in the near future.
In this section we also discuss briefly the expected
low-$Q^2$ behavior of the subtracted isoscalar polarization,
$\hat{\Pi}^{I=0}(Q^2)$, which can be obtained using values for the
relevant chiral LECs obtained from a chiral fit to the isovector
model data. Finally, in Sec.~\ref{sec4}, we discuss the relation between
the errors on the low-$Q^2$ contribution to $\hat{a}_\mu^{\rm LO,HVP}$
and those on the slope and curvature at $Q^2=0$, and argue that a
sub-percent determination of the former and few percent determination of the
latter should be sufficient to obtain a sub-percent determination of the
full contribution to $a_\mu^{\rm LO,HVP}$. This section also contains
our conclusions.

\begin{boldmath}
\section{\label{sec2}The model for $\hat{\Pi}^{I=1}(Q^2)$ and
its implications for the computation of $a_\mu^{\rm LO,HVP}$}

\subsection{\label{sec2a}A review of the model for $\hat{\Pi}^{I=1}(Q^2)$}
\end{boldmath}
The $I=1$ vector polarization function, $\Pi^{I=1}(Q^2)$, satisfies
a once-subtracted dispersion relation,
\begin{equation}
\label{disp}
\hat{\Pi}^{I=1}(Q^2)\, \equiv\, \Pi^{I=1}(Q^2)-\Pi^{I=1}(0)\,=\,-Q^2\, \int_{4m_\pi^2}^\infty ds\;
\frac{\rho (s)}{s(s+Q^2)}\ ,
\end{equation}
where $m_\pi$ is the pion mass, and $\rho (s)$ the corresponding spectral
function. A sensible choice for
$\Pi^{I=1}(0)$ and the function $\rho (s)$ thus determines a model for
$\Pi^{I=1}(Q^2)$.{\footnote{
$\Pi^{I=1}(0)$, of course, has no physical
significance, and is sensitive to the precise details of the short-distance
regularization of the two-point function.}}
The subtracted polarization represents one
such version, in which $\Pi^{I=1}(0)$ happens to be equal to $0$.

The spectral function $\rho (s)$ has been measured with high precision, for
$s<m_\tau^2$, in non-strange hadronic $\tau$ decays~\cite{alephud,opalud99}. In
Ref.~\cite{gmp13}, $\hat{\Pi}^{I=1}(Q^2)$ was determined from Eq.~(\ref{disp})
using as input a version of the OPAL data updated for modern values of the
exclusive mode branching fractions.{\footnote{Full details may be found
in the appendix of Ref.~\cite{dv72}.}} For those $s$ not accessible in
$\tau$ decay, $\rho (s)$ was represented by the 5-loop-truncated dimension
$D=0$ perturbative form \cite{PT}, supplemented by a model representation of the
residual, duality violating (DV) contribution. An exponentially damped
oscillatory form motivated by large-$N_c$
and Regge ideas, was used for the latter, based on a model for
duality violations
developed in Refs.~\cite{cgp08}, inspired by earlier work in Refs.~\cite{earlydv}.
Where the perturbative+DV form is used for $\rho (s)$ above $s=m_\tau^2$,
the DV contribution is much smaller than the perturbative one, making
the model dependence of the resulting version of $\hat{\Pi}^{I=1}(Q^2)$ extremely mild, especially in the low-$Q^2$ region where
the factor weighting $\rho (s)$, $1/[s(s+Q^2)]$, behaves as
$1/s^2$ over most of the spectrum. Our model for $\hat{\Pi}^{I=1}(Q^2)$
is thus a very physical one, especially so in the low-$Q^2$ region
most relevant to the $\hat{a}_\mu^{\rm LO,HVP}$ integral.
As such, it allows the systematics associated with various strategies
for the fitting of $\hat{\Pi}(Q^2)$ and evaluation of the integral
for $a_\mu^{\rm LO,HVP}$
to be investigated in a quantitative manner.
In taking the lessons from such model studies
over to the lattice, one must, of course, bear in mind that the value
of $\Pi^{I=1}(0)$ is not known on the lattice,
and will have to be determined either through a fit to the data or by
using time moments of the two-point function, as will be discussed
further below.\\

\begin{boldmath}
\subsection{\label{sec2b}Behavior
of the integrand of, and partial contributions to, $\hat{a}_\mu^{\rm LO,HVP}$}
\end{boldmath}

The physical model for $\hat{\Pi}^{I=1}(Q^2)$ described in the
previous section allows us to investigate in detail expectations, first,
for the behavior of the integrand in the $I=1$ analogue of Eq.~(\ref{amu})
and, second, for how rapidly (as a function of the upper limit of integration)
the contributions to $\hat{a}_\mu^{\rm LO,HVP}$ accumulate. To facilitate the
discussion below, we will denote by $\hat{a}_\mu^{\rm LO,HVP}[Q^2_{min},Q^2_{max}]$
the partial contribution to the $\hat{a}_\mu^{\rm LO,HVP}$ integral from the
interval $Q_{min}^2\le Q^2\le Q^2_{max}$. With this
notation, $\hat{a}_\mu^{\rm LO,HVP}[Q^2_{max}]=\hat{a}_\mu^{\rm LO,HVP}[0,Q^2_{max}]$ is the accumulated contribution
between $0$ and $Q^2_{max}$, and
$\hat{a}_\mu^{\rm LO,HVP}=\hat{a}_\mu^{\rm LO,HVP}[0,\infty ]$.

\begin{figure}[t]
\caption{\label{fig1}$f(Q^2)\, \hat{\Pi}^{I=1}(Q^2)$ versus $Q^2$
in the low-$Q^2$ region}
\centering
{\rotatebox{270}{\mbox{
\includegraphics[width=4.25in]
{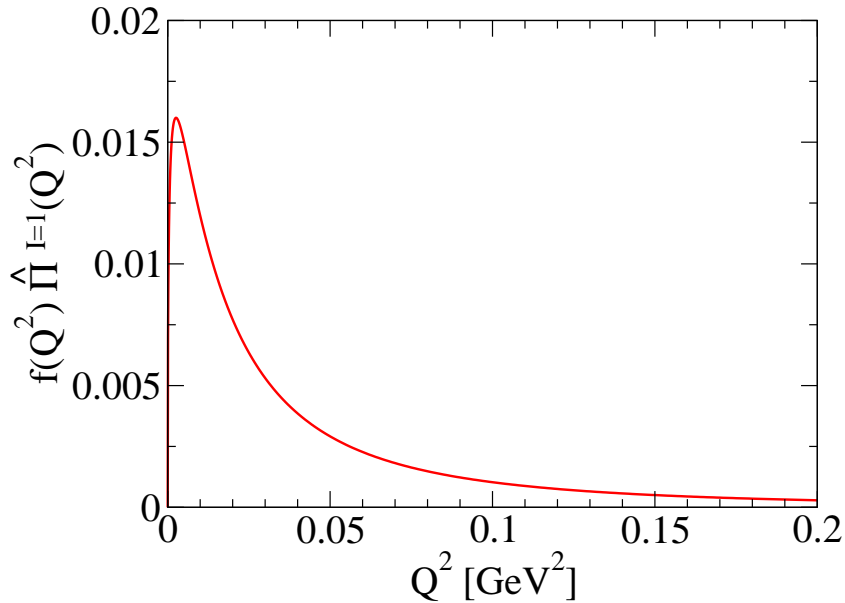}
}}}
\end{figure}

Figure~\ref{fig1} shows the product of the weight $f(Q^2)$ appearing
in the $a_\mu^{\rm LO,HVP}$ integral and the model version of the
subtracted $I=1$ polarization. As is well known,
this product is strongly peaked at low $Q^2$; it is thus shown
only in the region $Q^2<0.2$~GeV$^2$, beyond which it continues to decrease
rapidly and monotonically. The model shows the location of the peak to be
around $Q^2\sim m_\mu^2/4$. Lattice data typically does not reach such
low $Q^2$, and some form of fitting is thus necessary to extrapolate into
the peak region, at least in the conventional lattice approach.

\begin{figure}[t]
\caption{\label{fig2}The accumulation of the contributions to
$\hat{a}_\mu^{\rm LO,HVP}$ as a function of the upper
limit, $Q^2_{max}$, of integration. }
\centering
{\rotatebox{270}{\mbox{
\includegraphics[width=4.25in]
{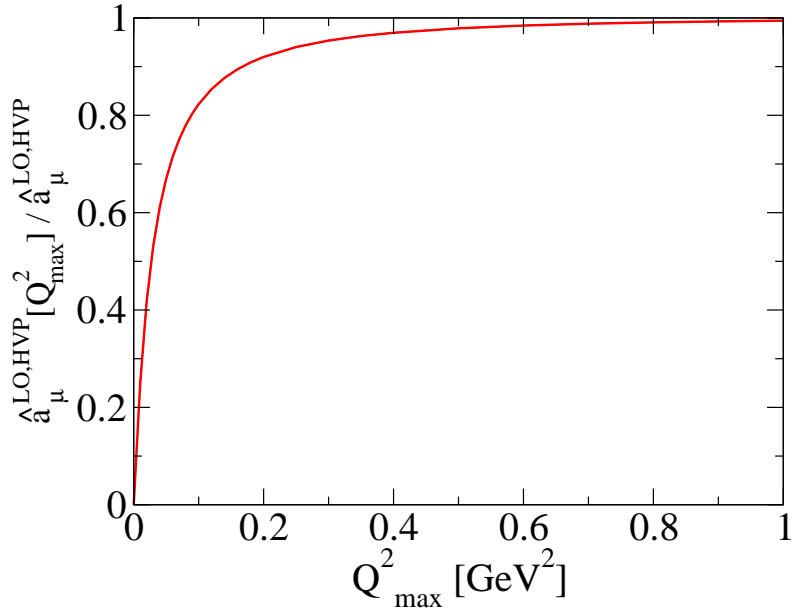}
}}}
\end{figure}

It is also useful to look at the accumulation of the contributions
to $\hat{a}_\mu^{\rm LO,HVP}$ as a function of the upper limit of
integration, $Q^2_{max}$. We display this accumulation, normalized to the
integral over all $Q^2$, $\hat{a}_\mu^{\rm LO,HVP}$, in the model,
in Fig.~\ref{fig2}.
We note that over $80\%$ of the contribution
is accumulated below $0.1$~GeV$^2$ and over $90\%$ below $0.2$~GeV$^2$.
It follows that the accuracy required for contributions above $0.1$ or
$0.2$~GeV$^2$ is much less than that required for the low-$Q^2$ region.
It thus becomes of interest to investigate the accuracy one might
achieve for the higher-$Q^2$ contributions were one to avoid
altogether fitting and/or modelling, and the associated systematic
uncertainty that accompanies it, and instead perform a direct
numerical integration over the lattice data. We investigate this
question in the next subsection.\\

\subsection{\label{sec2c}Direct numerical integration: how low can you go?}
In this section, we argue that existing lattice data, even
those without twisted boundary conditions, are already sufficiently accurate
that direct numerical integration of the lattice data can be relied
on to produce a value $\hat{a}_\mu^{\rm LO,HVP}[Q^2_{min},2$~GeV$^2]$ accurate
to well below $1\%$ of $\hat{a}_\mu^{\rm LO,HVP}$ for $Q^2_{min}$
down to about $ 0.1$~GeV$^2$. The situation will be even better once
the results of new data with reduced errors on $\Pi (Q^2)$ due
to all-mode averaging (AMA)~\cite{bis12,amaref} and/or denser sets of $Q^2$
produced by using twisted boundary conditions~\cite{DJJW12,ABGP13,HHJWDJ13} become available.

One practical issue, concerning the constant $\Pi^{I=1}(0)$ needed to convert the
unsubtracted polarization $\Pi^{I=1}(Q^2)$ obtained from the lattice to the
corresponding subtracted version $\hat{\Pi}^{I=1}(Q^2)$
needed for the $I=1$ analogue of the integral
in Eq.~(\ref{amu}), should be dealt with before continuing with the main
investigation of this section. The issue arises because the model we are
working with is one for the {\it subtracted} polarization. It thus
appears to differ from the lattice case, where a determination of $\Pi^{I=1}(0)$
and subsequent subtraction would be required.
This issue is, however, easily resolved. One simply
interprets the model, not as one for the subtracted polarization,
$\hat{\Pi}^{I=1}(Q^2)$, but rather as one for the unsubtracted polarization,
$\Pi^{I=1}(Q^2)$, happening to have $\Pi^{I=1}(0)=0$ and allows
$\Pi^{I=1}(0)$ to become a free parameter in fits of
data sets based on our model.{\footnote{Another
way of understanding what is going on here is as follows. The model
for the subtracted polarization can be converted to a related model more
closely resembling the lattice situation by simply adding a fixed constant
offset $C$ to all the subtracted polarization values
$\hat{\Pi}^{I=1}(Q^2)$. In
fitting fake data generated from this modified version of the model,
$\Pi^{I=1}(0)$ will of course need to be included as a fit parameter.
The result obtained for $\Pi^{I=1}(0)$ in such a fit will then
be exactly equal to the sum of $C$ and the result
$\hat{\Pi}^{I=1}(0)$ that would
be obtained by performing the same fit to the unmodified data with
$\hat{\Pi}^{I=1}(0)$} left free.}
The extent to which the fitted $\Pi^{I=1}(0)$ deviates from the known
value $0$ then quantifies the systematic uncertainty in the determination
of $\Pi^{I=1}(0)$ for the given fit function form.

Fits of $[1,1]$ and higher-order Pad\'e's on the interval between
$0$ and $1$~GeV$^2$ to the fake data set of Ref.~\cite{gmp13}
show that it is possible to obtain $\Pi^{I=1}(0)$ from such fits with an
uncertainty smaller than $0.001$.

\begin{figure}[t]
\caption{\label{fig3}The impact of an uncertainty
$\delta\Pi^{I=1}(0)=0.001$ in $\Pi^{I=1}(0)$ on
$\hat{a}_\mu^{\rm LO,HVP}[Q^2_{min},\infty]$
as a fraction of $\hat{a}_\mu^{\rm LO,HVP}$.}
\centering
{\rotatebox{270}{\mbox{
\includegraphics[width=4.25in]
{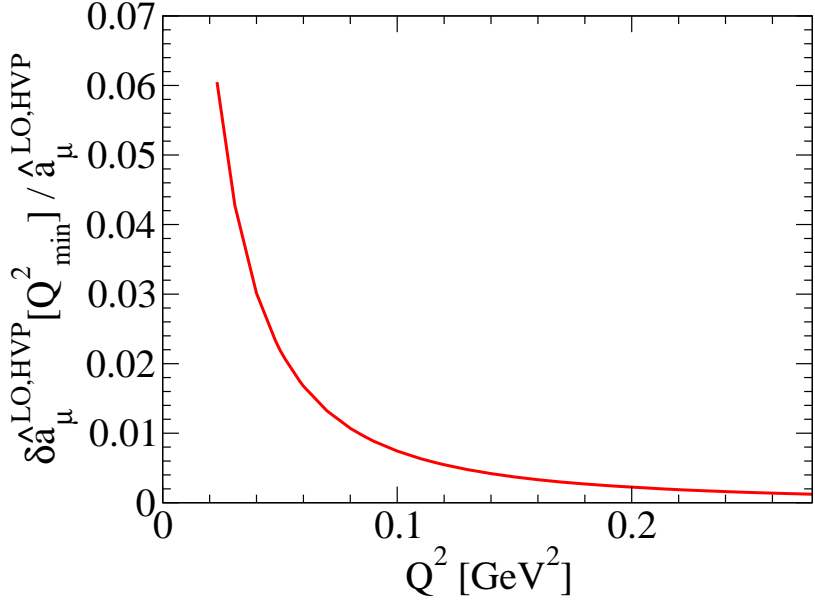}
}}}
\end{figure}

An uncertainty $\delta \Pi^{I=1}(0)$ produces a corresponding uncertainty
\begin{equation}
\delta\hat{a}_\mu^{\rm LO,HVP}[Q^2_{min}]\, =\,
4\alpha^2\, \delta\Pi^{I=1}(0)\, \int_{Q^2_{min}}^\infty dQ^2\,f(Q^2)
\end{equation}
on the contribution to $\hat{a}_\mu^{\rm LO,HVP}$ from
$Q^2\geq Q^2_{min}$. The
rapid decrease of $f(Q^2)$ with $Q^2$ means this uncertainty falls rapidly
with increasing $Q^2_{min}$. Figure~\ref{fig3} illustrates the impact of
this uncertainty on $\hat{a}^{\rm LO,HVP}$. The figure shows the $Q^2_{min}$
dependence of $\delta\hat{a}^{\rm LO,HVP}[Q^2_{min},\infty ]$,
as a fraction of $\hat{a}^{\rm LO,HVP}$, for $\delta\Pi^{I=1}(0)=0.001$.
Even with this (what we expect to be rather conservative) choice for
$\delta\Pi^{I=1}(0)$, the error remains safely below $1\%$ for
$Q^2_{min}$ down to $0.1$~GeV$^2$, where
\begin{equation}
\frac{\delta\hat{a}_\mu^{\rm LO,HVP}[0.1\ \mbox{GeV}^2]}
{\hat{a}_\mu^{\rm LO,HVP}}\, =\,
0.0074\, \left({\frac{\delta\Pi^{I=1}(0)}{0.001}}\right)\ .
\end{equation}
The relatively rapid growth
at lower $Q^2_{min}$, however, means that careful monitoring of this error
for the $\delta\Pi^{I=1}(0)$
actually achieved in a given analysis would be
required if one wished to push the lower limit of direct numerical
integration of the lattice data to below $0.1$~GeV$^2$.

We now turn to the model study of the accuracy of the direct numerical
integration of the subtracted polarization data, assuming that
$\delta\Pi^{I=1}(0)$ is small enough to allow for a sufficiently precise subtraction. For this purpose, we
employ the fake $I=1$ data set used previously in Ref.~\cite{gmp13}.
The set was constructed from the $\tau$-data-based model discussed above
using the covariance matrix for a $64^3\times 144$ MILC ensemble with
periodic boundary conditions, $a\approx 0.06$~fm and $m_\pi\approx 220$~MeV
\cite{MILC}.

The lattice covariance matrix is, by construction, also the covariance matrix
of the fake data set. With the fake data and its covariances in hand, we
evaluate $\hat{a}_\mu^{\rm LO,HVP}[Q^2_{min},2$~GeV$^2]$ and its error by direct
trapezoid rule integration of the data, and compare the result to the
corresponding exact result in the model. The difference between the two
gives the systematic error associated with estimating
$\hat{a}_\mu^{\rm LO,HVP}[Q^2_{min},2$~GeV$^2]$ by direct numerical integration.\footnote{The choice $Q^2_{max}=2$~GeV$^2$ is somewhat arbitrary, but
in our model $\hat{a}_\mu^{\rm LO,HVP}[2$~GeV$^2]$ is 99.74\% of $\hat{a}_\mu^{\rm LO,HVP}$.}

In addition to this systematic uncertainty, there is, of course, also
the statistical component of the overall uncertainty, obtained by propagating
the data covariances through the trapezoid-rule evaluation. In the present
model study, these covariances are those of the fake data set.

\begin{figure}[t]
\caption{\label{fig4}The systematic and statistical components of the
error on the evaluation of $\hat{a}_\mu^{\rm LO,HVP}[Q^2_{min},2$~GeV$^2]$
by direct trapezoid-rule numerical integration, as a fraction of
$\hat{a}_\mu^{\rm LO,HVP}$.}
\centering
{\rotatebox{270}{\mbox{
\includegraphics[width=4.25in]
{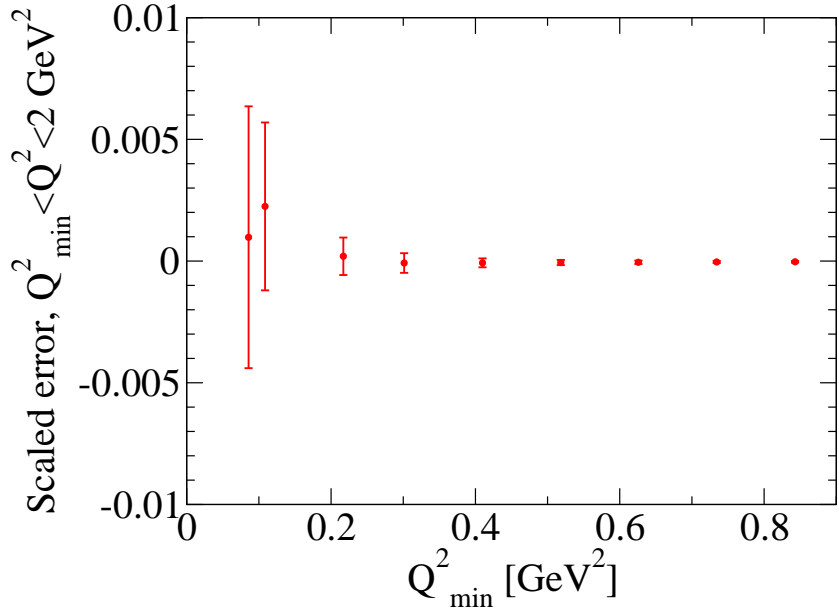}
}}}
\end{figure}

The results for both the systematic and statistical components of the
uncertainty on the trapezoid rule evaluation are displayed, as a function of
$Q^2_{min}$, in Fig.~\ref{fig4}. For each $Q^2_{min}$, the
displayed central value represents the corresponding systematic uncertainty,
while the error bar gives the size of the corresponding statistical
uncertainty. The results have been scaled by $\hat{a}_\mu^{\rm LO,HVP}$ in
order to display the impact of the numerical integration
uncertainty on the final error for $\hat{a}_\mu^{\rm LO,HVP}$. We see that
both components are completely negligible above $Q^2_{min}\approx 0.2$~GeV$^2$.
The systematic component remains below $0.25\%$ for all points shown. The statistical component is seen to be dominant
for low $Q^2_{min}$, reaching about $0.5\%$ for the lowest value shown
($Q^2_{min}=0.086$~GeV$^2$). The growth of the statistical component
with decreasing $Q^2_{min}$ is a consequence of the rapid growth in the
data errors for the very low-$Q^2$ points, something that would be
significantly reduced with improved data \cite{bis12,amaref}.

The results of this study show that data from existing lattice simulations,
even without twisted boundary conditions and/or AMA improvement, allow an
evaluation of the contributions to $\hat{a}^{\rm LO,HVP}$ from $Q^2>Q^2_{min}$
with an accuracy safely below $1\%$ of $\hat{a}^{\rm LO,HVP}$ for $Q^2_{min}$
down to at least $0.1$~GeV$^2$. While not yet available, analogous fake data
sets constructed from covariance matrices corresponding to lattice data
with twisted boundary conditions and AMA improvement,
will, once available, allow us to quantify the level of
improvement made possibly by better statistics and a
finer distribution of $Q^2$ points.  Of course,
as explained at the beginning of this subsection, $\Pi^{I=1}(0)$,
needed to compute $\hat{\Pi}^{I=1}(Q^2)$
for the numerical integration, will have to be
determined with sufficient precision as well.

The fact that $\hat{a}_\mu^{\rm LO,HVP}[Q^2_{min},2$~GeV$^2]$ can be reliably
evaluated by direct numerical integration down to $Q^2_{min}\sim 0.1$~GeV$^2$
greatly simplifies the task of computing the rest of the
contribution to $\hat{a}_\mu^{\rm LO,HVP}$. The reason is that,
for $0\leq Q^2\,\ltap\, 0.1$~GeV$^2$, one expects fits using
low-order Pad\'e's of the types proposed in Refs.~\cite{abgp12,hpqcd14},
or using the conformal polynomial or chiral representations discussed below
(Secs. \ref{sec3b} and \ref{sec3c}),
to provide efficient and reliable representations of the subtracted
polarization function. We show that this is indeed the case in the next
section, and investigate the systematic uncertainties
on the low-$Q^2$ contributions produced by the use of such fit forms.

\begin{boldmath}
\section{\label{sec3}Behavior of the subtracted polarization in
the low-$Q^2$ region and a hybrid strategy for evaluating $a_\mu^{\rm LO,HVP}$}
\end{boldmath}

In the previous section we showed that contributions to
$\hat{a}_\mu^{\rm LO,HVP}$ from $Q^2$ above $\sim 0.1$~GeV$^2$ can
be obtained with an accuracy better than $1\%$ of
$\hat{a}_\mu^{\rm LO,HVP}$ by direct numerical
integration of existing lattice data. In this section, we discuss
the region between $0$ and $\sim 0.1$~GeV$^2$ and
investigate the reliability of low-order Pad\'e, conformally mapped
polynomial, and ChPT representations
of the subtracted polarization in this region.  We focus
on the systematic accuracy achievable using these representations for the evaluation
of the low-$Q^2$ contributions to $\hat{a}_\mu^{\rm LO,HVP}$.
As in the previous
sections, these investigations are performed using the $\tau$-data-based
model for $\hat{\Pi}^{I=1}(Q^2)$.

At low $Q^2$, fits of lattice data to a functional form
are needed to achieve a precise determination of the integral
in Eq.~(\ref{amu}). To avoid difficult-to-quantify
systematic errors, the form(s) employed should be
free of model dependence.  Here we investigate three such functional
forms, one based on a sequence of Pad\'e approximants \cite{abgp12,hpqcd14},
one based on a conformally mapped polynomial, and one
based on ChPT.   An important question is to what order
Pad\'e, what degree conformally mapped polynomial, and what order
in the chiral counting one must go in order to obtain
representations of $\hat{\Pi}^{I=1}(Q^2)$ of sufficient accuracy.
In addition, there is the question of
with what statistical precision these functional forms can
then be fit to lattice data. Even if
in principle a certain functional form provides an accurate representation of
$\hat{\Pi}^{I=1}(Q^2)$, the parameters still have to be determined with sufficient
precision.   In this article, we address only the first question, leaving an
investigation of the second question to the future, when much more precise
lattice data at low $Q^2$ are expected to become available.

In order to probe the accuracy of an approximate functional form in
representing the exact function $\hat{\Pi}^{I=1}(Q^2)$, we need to fix
the parameters of that form. We will do so
by constructing the Pad\'e, conformal and chiral
representations such that they reproduce the values
of the the relevant low-order derivatives of
$\hat{\Pi}^{I=1}(Q^2)$ with respect to $Q^2$
at $Q^2=0$. In the model case, these derivatives
are known from the dispersive representation of the
subtracted polarization, while on the lattice they can be obtained
from time moments of the vector current two-point function, as explained
in more detail below.  Since we are concerned with the
systematic uncertainty associated with the use of a given functional
form in the low-$Q^2$ region, we will assume these derivatives to
be exactly known and given by the central values resulting from
the dispersive representation. It will still be necessary to reduce the
errors on the low-$Q^2$ lattice data in order to bring the corresponding
statistical uncertainties under control. Our goal is thus only to
identify those functional forms which produce systematic uncertainties
at the sub-percent
level when used with future improved low-$Q^2$ data.\\

\subsection{\label{sec3a}Low-order Pad\'e representations of
the subtracted polarization}
As already pointed out in Ref.~\cite{abgp12},
the function $\Phi (Q^2)\equiv -\hat{\Pi}^{I=1}(Q^2)/Q^2$ is a so-called
Stieltjes function and, as such, satisfies a number of theorems on convergent
representations over compact regions of the complex $Q^2$ plane via Pad\'e
approximants~\cite{multiptpades,oneptpades}. For example,
the sequence of $[M+J,M]$ Pad\'e's constructed to match the first
$N=2M+J+1$ coefficients of the Taylor expansion of $\Phi (Q^2)$ about
$Q^2=0$ is known to converge to $\Phi (Q^2)$ as $M\rightarrow\infty$, and
for any $J\geq -1$, in any compact set in the complex $Q^2$-plane not
overlapping the cut of $\hat{\Pi}^{I=1}$~\cite{oneptpades}. Moreover,
for $Q^2>0$, the
set of such Pad\'e's satisfies the inequalities~\cite{oneptpades}
\begin{eqnarray}
&&[0,1]\leq [1,2]\leq \cdots \leq [N,N+1]\leq \Phi (Q^2)
\leq [N,N]\leq \cdots \leq [1,1]\leq [0,0]\ .\label{padeineq}
\end{eqnarray}
To make contact with the notation employed in Ref.~\cite{hpqcd14},
let us denote $-Q^2$ times the $[M,N]$ Pad\'e in (\ref{padeineq})
by $[M+1,N]_H$. The inequalities (\ref{padeineq}) then
correspond to the following inequalities for the Pad\'e representations of
$\hat{\Pi}^{I=1}(Q^2)$
\begin{eqnarray}
&&[1,0]_H\leq [2,1]_H\leq \cdots \leq [N+1,N]_H\leq
\hat{\Pi}^{I=1}(Q^2) \nonumber\\
&&\qquad\qquad \leq [N,N]_H\leq \cdots
\leq [2,2]_H \leq [1,1]_H\ .\label{padeineqpr}
\end{eqnarray}

In Ref.~\cite{hpqcd14} it has been pointed out that the derivatives
of the polarization function at $Q^2=0$, needed to construct the sequences of
Pad\'e's in Eq.~(\ref{padeineqpr}), can be determined by evaluating even-order
Euclidean time moments of the zero-spatial-momentum
representation of the relevant vector current two-point function on the
lattice.\footnote{For an alternative approach to obtaining $\Pi(0)$,
see Ref.~\cite{DPT12}. }
This idea was implemented for the $\bar{s}s$ and $\bar{c}c$
vector current polarization functions and the resulting representations
used to determine the strange and charm contributions to $a_\mu^{\rm LO,HVP}$.
Evidence was presented that convergence has been achieved by the time
the $[2,2]_H$ order is reached. However, in the light-quark sector
the errors on these moments are expected to be much larger, and to
grow rapidly with increasing order, because light-quark
correlators are very noisy at large Euclidean $t$.
It is, first of all, not clear
what order Pad\'e would be required for suitable convergence in the
light-quark sector and, second, not obvious that the moments needed
to construct, {\it e.g.}, the $[2,2]_H$ Pad\'e can be determined with
sufficient accuracy to make the computation of the full light-quark
contribution to $a_\mu^{\rm LO,HVP}$ feasible in this approach.

\begin{figure}[t]
\centering
{\rotatebox{270}{\mbox{
\includegraphics[width=3.5in]{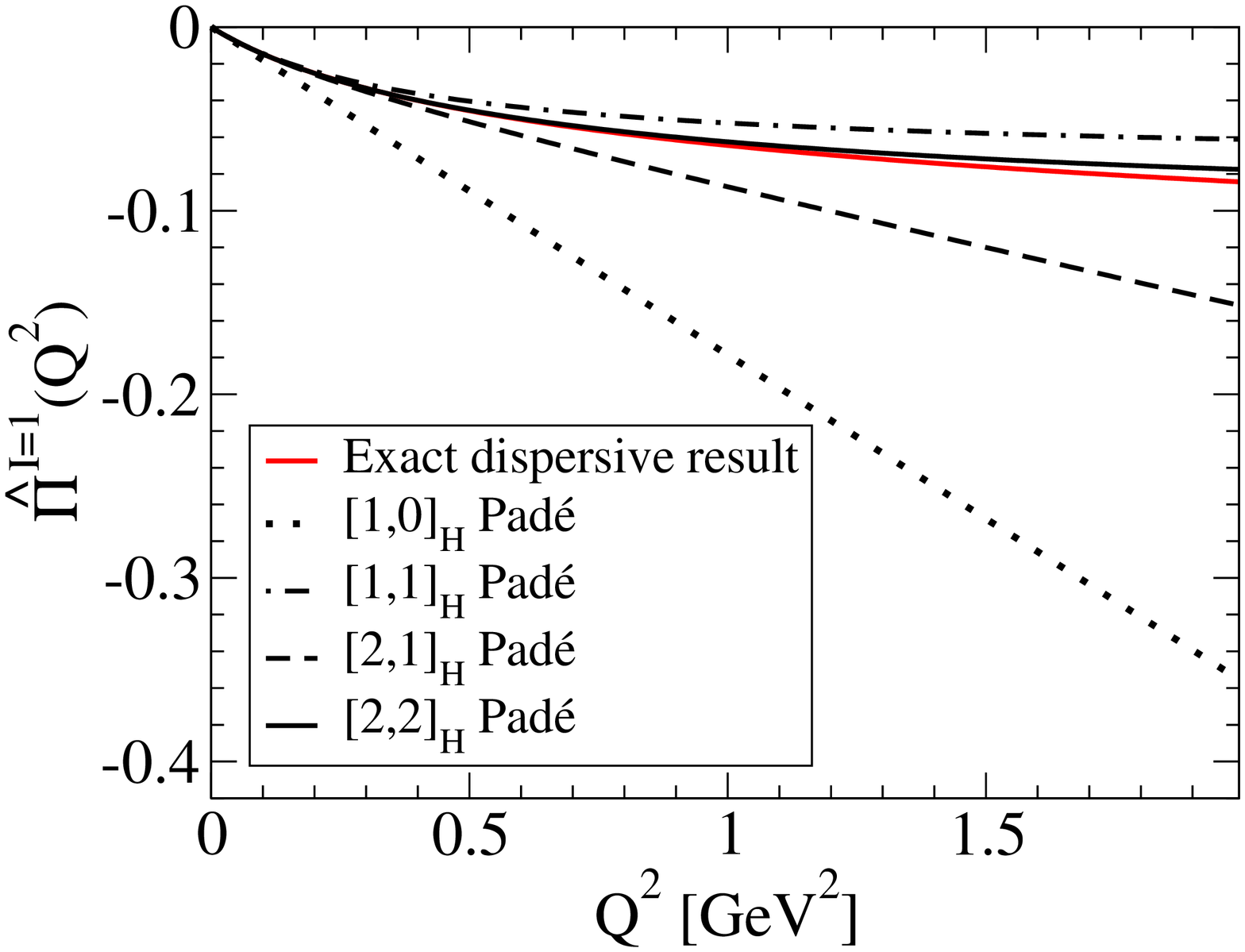}
}}}
\hspace{.1cm}
{\rotatebox{270}{\mbox{
\includegraphics[width=3.5in]{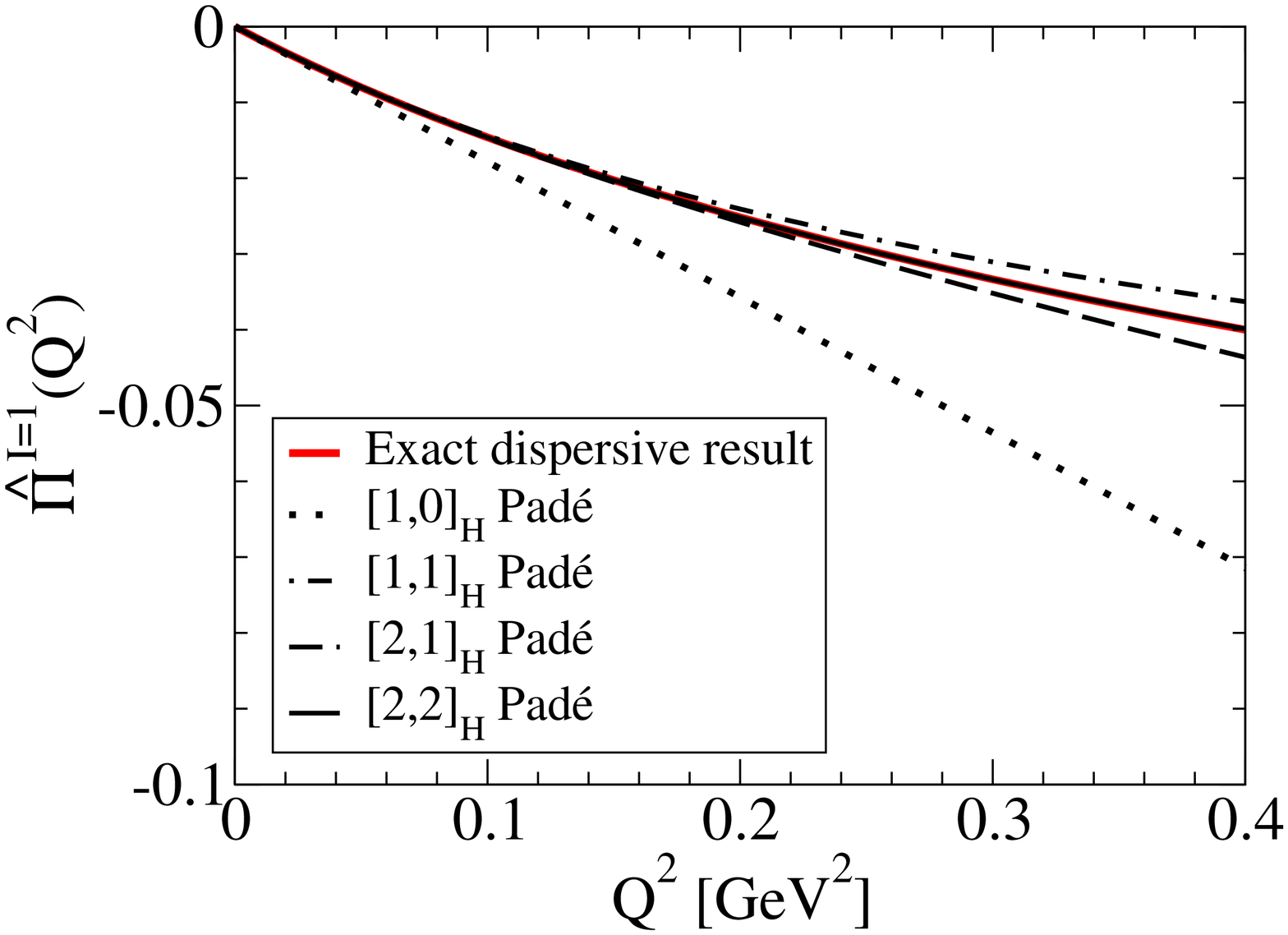}
}}}
\caption{\label{fig5}Comparison of the exact dispersive model results for
$\hat{\Pi}^{I=1}(Q^2)$ with the Pad\'e's constructed from the
derivatives of the model with respect to $Q^2$ at $Q^2=0$ in
the intervals $0\le Q^2\le 2$~GeV$^2$ (upper panel) and $0\le Q^2\le 0.4$~GeV$^2$
(lower panel).}
\vspace*{2ex}
\end{figure}

The $\tau$-data-based model for $\hat{\Pi}^{I=1}(Q^2)$ provides a convenient
tool for investigating the first of these questions. First, since the
exact values of the derivatives of $\hat{\Pi}^{I=1}(Q^2)$ with
respect to $Q^2$ at $Q^2=0$ in the model are easily obtained
from the dispersive
representation, Eq.~(\ref{disp}), it is straightforward to construct
the exact-model versions of the Pad\'e's of Ref.~\cite{hpqcd14}
and see how well they do in representing $\hat{\Pi}^{I=1}(Q^2)$.
Second, knowing that contributions to $\hat{a}^{\rm LO,HVP}$
from $Q^2$ above $\sim 0.1$~GeV$^2$ can be accurately determined by
direct numerical integration of existing lattice data, we can use
the model to explore the obvious question raised by this observation,
namely how low an order of Pad\'e will suffice if one's goal is to
evaluate the contribution to $\hat{a}_\mu^{\rm LO,HVP}$,
not for all $Q^2$,
but rather only for the restricted region $0\le Q^2\,\ltap\, 0.1$~GeV$^2$.

Figure~\ref{fig5} shows the comparison of the dispersive results for
$\hat{\Pi}^{I=1}(Q^2)$ and the $[1,0]_H$, $[1,1]_H$, $[2,1]_H$,
and $[2,2]_H$ Pad\'e's constructed using the exact dispersive results
for the derivatives of $\hat{\Pi}^{I=1}(Q^2)$ with respect to $Q^2$ at
$Q^2=0$. The top panel shows the comparison in the inteval
$0\le Q^2\le 2$~GeV$^2$, the bottom panel the same comparison in the more
restricted region $0\le Q^2\le 0.4 $~GeV$^2$.   Note that the curves
shown in this figure  follow the pattern of the inequalities
in Eq.~(\ref{padeineqpr}).
We see that the $[2,2]_H$
Pad\'e provides a good, though not perfect, representation of
$\hat{\Pi}^{I=1}(Q^2)$ over the whole of the range $0\le Q^2\le 2$~GeV$^2$.
This is not true of the lower-order Pad\'e's. When one focuses on
the low-$Q^2$ region, however, it is evident that even the $[1,1]_H$
Pad\'e provides a very accurate representation in the region of current
interest, $0\le Q^2\le 0.2$~GeV$^2$.

\begin{figure}[t]
\centering
{\rotatebox{270}{\mbox{
\includegraphics[width=3.5in]
{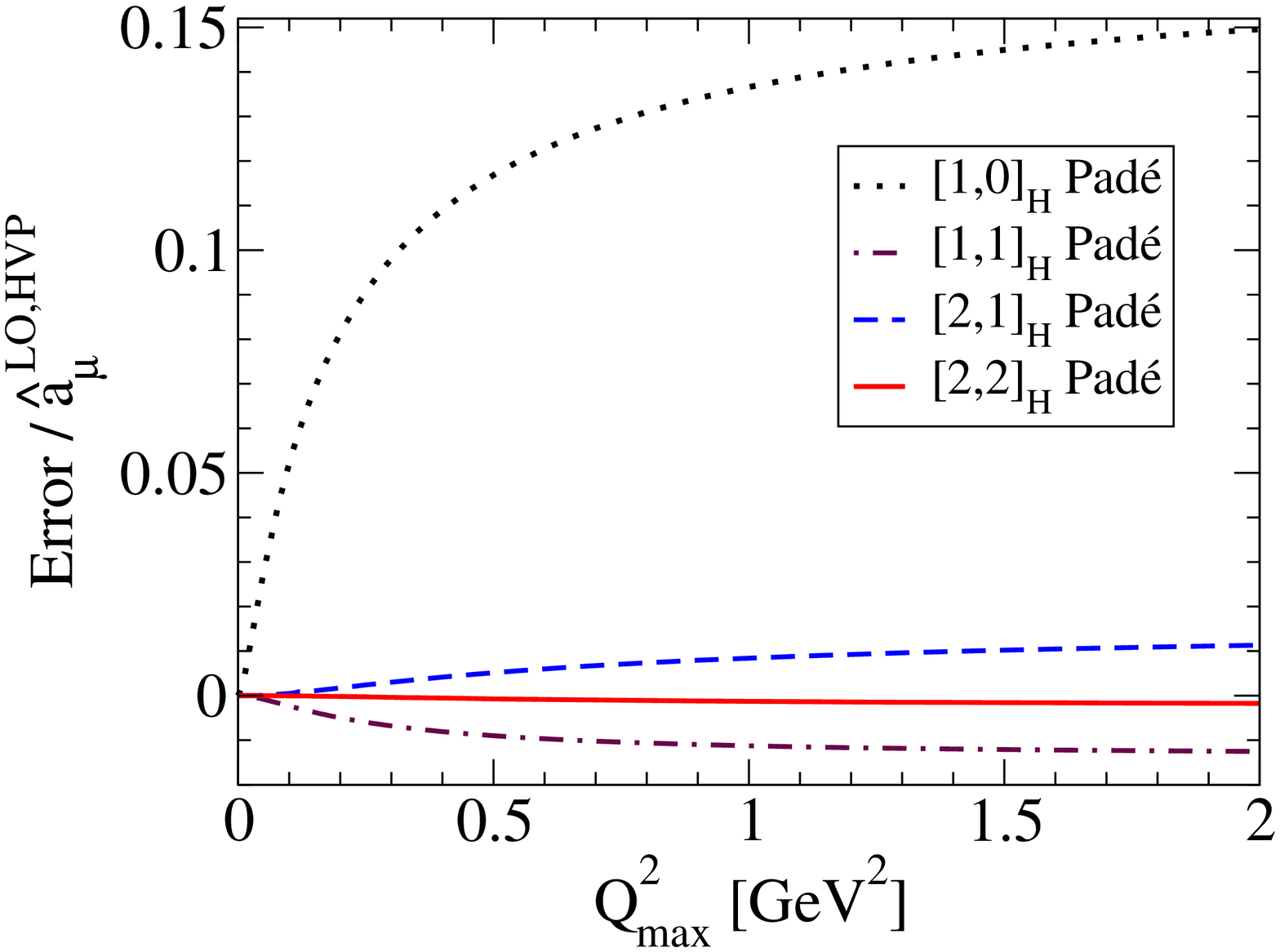}
}}}
\hspace{.1cm}
{\rotatebox{270}{\mbox{
\includegraphics[width=3.5in]
{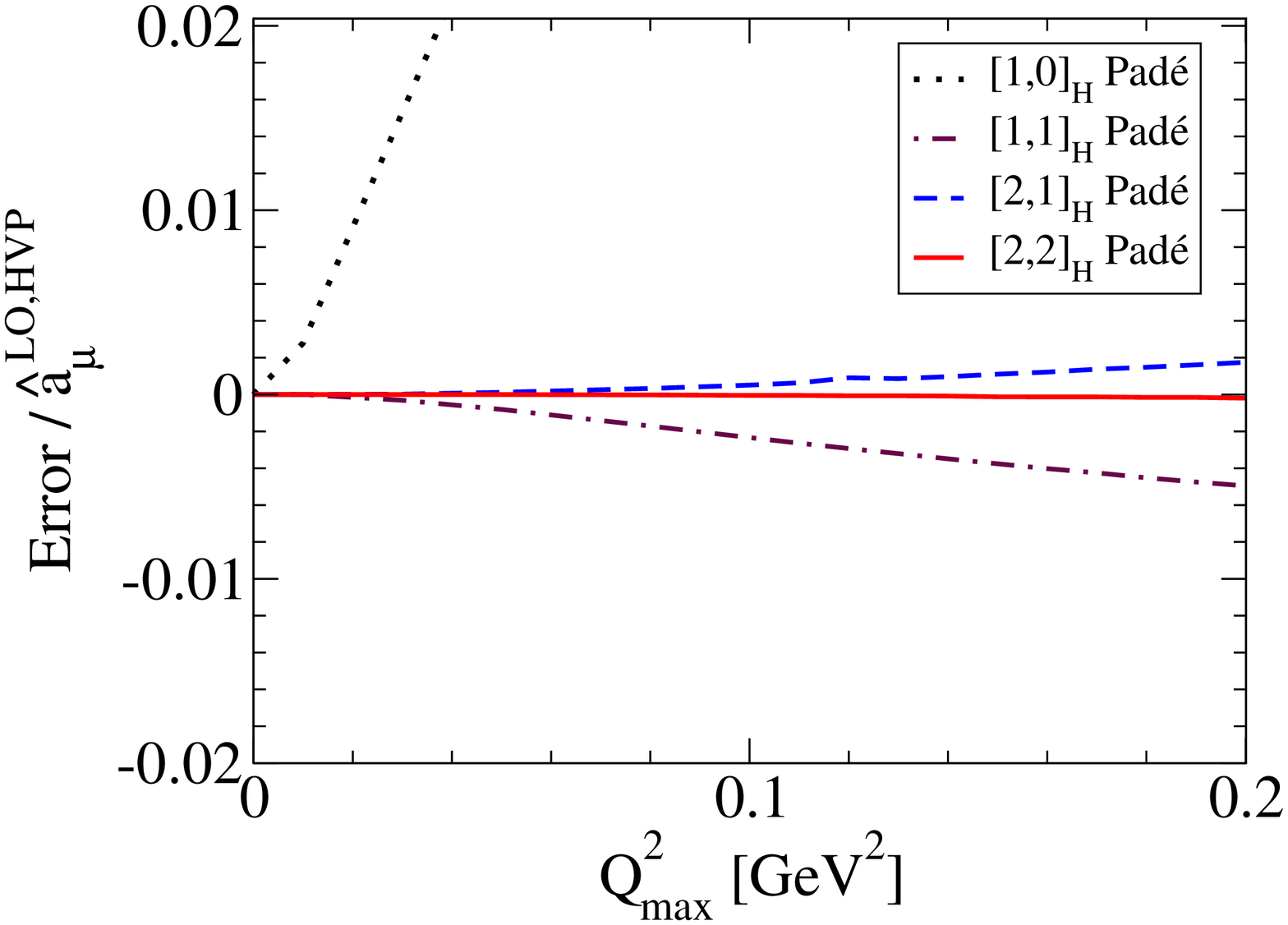}
}}}
\caption{\label{fig6}Deviations of the Pad\'e estimates for
$\hat{a}_\mu^{\rm LO,HVP}[Q^2_{max}]$ as a fraction of
$\hat{a}_\mu^{\rm LO,HVP}$ in the intervals
$0\le Q^2_{max}\le 2$~GeV$^2$ (upper panel) and $0\le Q^2_{max}\le 0.2$~GeV$^2$
(lower panel).  Note the difference in scale on the vertical axis.
}
\vspace*{2ex}
\end{figure}

For the problem at hand, of course, it is deviations of the Pad\'e
representations from $\hat{\Pi}^{I=1}(Q^2)$ in the low-$Q^2$
region that are of importance in determining the accuracy
of the Pad\'e-based estimates for $\hat{a}_\mu^{\rm LO,HVP}$. The impact of
the deviations seen in Fig.~\ref{fig5} on the contribution
$\hat{a}_\mu^{\rm LO,HVP}[Q^2_{max}]$ from the region
$0\le Q^2\le Q^2_{max}$ is shown
in Fig.~\ref{fig6} as a function of $Q^2_{max}$. The upper panel shows
the difference between the various order Pad\'e estimates and the exact
model result, scaled as usual by $\hat{a}_\mu^{\rm LO,HVP}$, for $Q^2_{max}$
in the interval $0\le Q^2_{max}\le 2$~GeV$^2$, while the lower panel zooms
in on the region below $0.2$~GeV$^2$ of interest here.

We see that, if one insists on using the time moments to evaluate the
contributions to $\hat{a}_\mu^{\rm LO,HVP}$ from $Q^2$ out to
$Q^2_{max}=2$~GeV$^2$
or above, reducing the systematic error on the evaluation to below $1\%$
will require going to the $[2,2]_H$ Pad\'e.
This would necessitate evaluating time moments with good accuracy out to
tenth order, which is likely to be a challenging task for light-quark
two-point functions.

We have seen, however, that there is no need to push the moment-based evaluation
of $\hat{a}^{\rm LO,HVP}_\mu[Q^2_{max}]$
out to $Q^2_{max}\sim 2$~GeV$^2$. In the region below $Q^2\sim 0.1-0.2$~GeV$^2$
which cannot be handled by direct numerical integration of the lattice
data, one does not need the $[2,2]_H$ Pad\'e to achieve an accurate
representation of $\hat{\Pi}^{I=1}(Q^2)$. The lower panel of Fig.~\ref{fig6}
shows that even the $[1,1]_H$ representation is sufficient
in this region, producing an estimate for
$\hat{a}_\mu^{\rm LO,HVP}[Q^2_{max}]$ accurate to about 0.3\% for
$Q^2_{max}=0.1$~GeV$^2$ and to about $0.5\%$ even for
$Q^2_{max}=0.2$~GeV$^2$. This is a potentially
significant advantage since constructing the $[1,1]_H$ Pad\'e
requires moments only up to sixth order.
The $[2,1]_H$ Pad\'e lowers the previous errors to
$0.06\%$ and $0.2\%$, respectively,
but it requires the eighth order moment in its construction.

It is worth emphasizing that another sequence of Pad\'e approximants to
$\Pi^{I=1}(Q^2)$ exists; these are the multi-point Pad\'e's of
Ref.~\cite{abgp12}, for which convergence theorems also
exist~\cite{multiptpades}.  These multi-point Pad\'e's actually have the
same form as the single-point, $Q^2=0$ Pad\'e's discussed in
Ref.~\cite{hpqcd14}.{\footnote{Refs.~\cite{abgp12} and \cite{hpqcd14},
unfortunately, use different notations to specify what end
up being the same Pad\'e representation of $\hat{\Pi}(Q^2)$. The
Pad\'e denoted $[M,N]$ in Ref.~\cite{abgp12} corresponds to what is
called $[M+1,N]$ in Ref.~\cite{hpqcd14}. We employ the alternate
notation $[M+1,N]_H$, introduced already above, for the latter
in order to distinguish between it and the earlier notation
employed in Ref.~\cite{abgp12}.}}
Fitting the coefficients of
such Pad\'e's over a relatively low-$Q^2$ interval in which the Pad\'e in
question is known to provide an accurate representation of
$\hat{\Pi}^{I=1}(Q^2)$ is thus an alternative to obtaining these
coefficients by evaluating the time moments of the two-point function.
Which of the two approaches will yield the smallest
statistical error is a topic
for future investigation.

One should, however, bear in mind in this regard that the time moments,
in producing the derivatives of the subtracted polarization with respect
to $Q^2$ at $Q^2=0$, will yield Pad\'e's which, by construction, will be most
accurate in the low-$Q^2$ region of primary interest for evaluating
$\hat{a}_\mu^{\rm LO,HVP}$. The deviations of the Pad\'e
constructed in this
manner from the underlying subtracted polarization will thus lie at
higher $Q^2$ and have a reduced impact on the error on
$\hat{a}_\mu^{\rm LO,HVP}$, if the Pad\'e is only
used to get the low-$Q^2$ contribution.
In contrast, in fitting the coefficients of the Pad\'e's using low-$Q^2$
data, the fits will inevitably be more heavily constrained by the somewhat
larger $Q^2$ points in the fit interval, as these will have smaller errors
than the points at very low $Q^2$. The resulting Pad\'e may thus
be less accurate at very low $Q^2$, and one may need to go to a
higher-order Pad\'e in comparison to the moment-based approach.
Further quantitative investigations of the lattice situation
will become possible
once covariance matrices corresponding to lattice data with twisted boundary
conditions and AMA improvement become available.

For now, however,
we can investigate this issue only using the $I=1$ model data and
associated covariance matrix, the latter being generated by the
covariances of the experimental $\tau$-decay data used in constructing
the model.
We emphasize that this covariance matrix is very different from what we
may expect any covariance matrix coming from lattice data to look like,
so the following short exercise can serve only to address systematic
issues, and has nothing to say about the statistics that will be
required on the lattice.

If we fit the $\tau$-based data on the interval between $0.1$ and
$0.2$~GeV$^2$,
where lattice errors will typically be much smaller
than those at lower $Q^2$,
we find that it is necessary to go to the
$[2,1]_H$ Pad\'e if one wishes to reduce the systematic uncertainty
on the low-$Q^2$ Pad\'e determination of
$\hat{a}_\mu^{\rm LO,HVP} [0.1$~GeV$^2]$ to the sub-percent level.
  As an example, a fit to model data at the
points $Q^2=0.10,\, 0.11,\, \cdots ,\, 0.20$~GeV$^2$ using the
$[2,1]_H$ Pad\'e form, with $\hat{\Pi}^{I=1}(0)$ a free parameter,
yields an estimate for $\hat{a}_\mu^{\rm LO,HVP}[0.1$~GeV$^2]$ accurate
to better than $0.3\%$ of $\hat{a}_\mu^{\rm LO,HVP}$. Even more useful,
though not unexpected in view of the fact that the $[2,1]_H$
representation is essentially indistinguishable from the underlying
model polarization out to $Q^2\approx 0.2$~GeV$^2$,
$\hat{a}_\mu^{\rm LO,HVP}[Q^2_{max}]$ remains accurate to better than $0.3\%$
out to $Q^2_{max}=0.2$~GeV$^2$. This means that, with sufficiently
good data in the interval between $Q^2\approx 0.1$ and $0.2$~GeV$^2$,
one would be able to vary the choice of boundary $Q_{min}^2$
between the low-$Q^2$ and high-$Q^2$ regions and obtain
combined hybrid determinations of the full contribution to
$a_\mu^{\rm LO,HVP}$ for several choices of $Q_{min}^2$,
providing further checks on the systematics of
the hybrid approach.

\subsection{\label{sec3b}Conformal expansion of the subtracted
polarization}
The Taylor expansion
of $\Pi^{I=1}(Q^2)$ in the variable $Q^2$ converges
for $\vert Q^2\vert <4m_\pi^2$.  However, with $4m_\pi^2=0.078$~GeV$^2$,
the radius of convergence is most likely too small to be useful in practice.
We can improve the convergence properties by rewriting
$\Pi^{I=1}(Q^2)$ first in terms of the variable
\begin{equation}
\label{exppar}
w(Q^2)=\frac{1-\sqrt{1+z}}{1+\sqrt{1+z}}\ ,\qquad z=\frac{Q^2}{4m_\pi^2}\ ,
\end{equation}
and then expanding in $w$.  The series
\begin{equation}
\label{series}
\Pi^{I=1}(Q^2)=\sum_{n=0}^\infty p_n w^n
\end{equation}
should have better convergence properties than the Taylor expansion in
$z$, because the whole complex $z$ plane is mapped onto
the unit disc in the complex $w$ plane, with the cut $z\in(-\infty,-1]$
mapped onto the disc boundary. The expansion~(\ref{series}) thus
has radius of convergence $\vert w\vert =1$.
In terms of the variable $Q^2$, this includes the positive real axis.

\begin{figure}[t]
\caption{\label{fig7}Comparison of the results of the conformal
polynomial
representations up to quadratic order with the exact
$\tau$-data-based model  for  $\hat{\Pi}^{I=1}(Q^2)$. }
\centering
{\rotatebox{270}{\mbox{
\includegraphics[width=4.25in]
{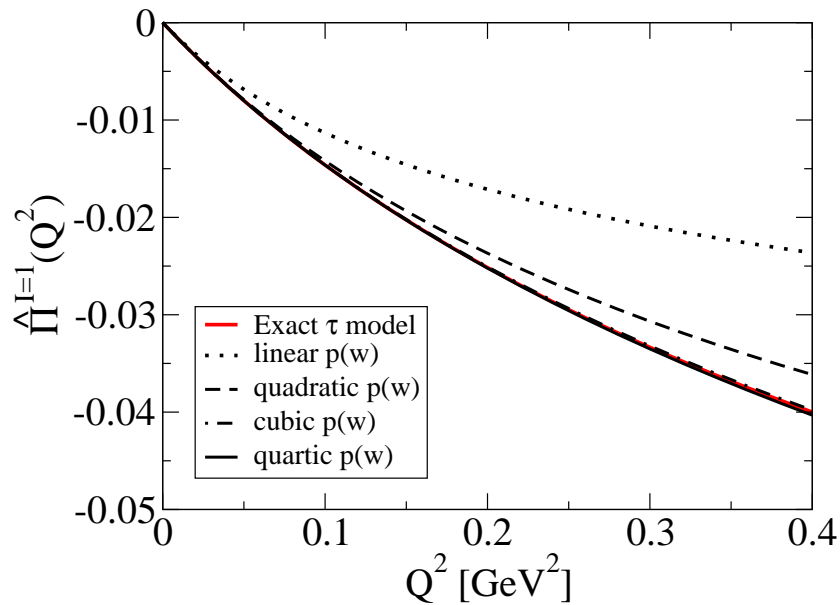}
}}}
\end{figure}

For the coefficients $p_1,\, p_2,\, \cdots ,\, p_4$ needed to
construct $p(w)$ up to degree $4$, we find, from the derivatives of
$\Pi^{I=1}(Q^2)$ with respect to $Q^2$ at $Q^2=0$ in the model, the
values $p_1=0.05565$ and $p_2\, =\, -0.06936$, $p_3=0.04781$ and
$p_4\, =\, -0.01561$. The resulting representations of $\hat{\Pi}^{I=1}(Q^2)$
linear, quadratic, cubic and quartic in $w$ are compared to the exact
model values in Fig.~\ref{fig7}.
We observe, from Figs.~\ref{fig5} and \ref{fig7}, that the
Pad\'e and conformal polynomial representations with the same number
of parameters lie close to one another.

Let us look more closely at the values of
$\hat{a}_\mu^{\rm LO,HVP}[Q^2_{max}]$ obtained from the conformal
polynomial representations. The quadratic version, for example,
yields estimates for $\hat{a}_\mu^{\rm LO,HVP}[Q^2_{max}]$
$0.6\%$ and $1\%$ below the exact model values for $Q^2_{max}=0.1$ and
$0.2$~GeV$^2$, respectively, while the corresponding errors for the
cubic representation are $0.02\%$ and $0.04\%$.  These numbers are to
be compared to $0.3\%$ and $0.5\%$ for the $[1,1]_H$ Pad\'e
(which has the same number of parameters as the quadratic polynomial),
and $0.06\%$ and $0.2\%$ for the $[2,1]_H$ Pad\'e (which has
same number of parameters as the cubic polynomial).

While the higher-order conformal representations discussed above
provide very accurate results for $\hat{a}_\mu^{\rm LO,HVP}[Q^2_{max}]$,
one should bear in mind that their construction requires as input the
values of the derivatives of $\hat{\Pi}(Q^2)$ with respect to $Q^2$ at
$Q^2=0$.  As mentioned before, these can, in principle,
be obtained from the time moments
of the two-point function. Accurate determinations of the relevant
moments will thus be required to make the conformal approach useful
in this form. It is, of course, also possible to implement
the conformal representation by fitting the coefficients of a truncated
version of the expansion in Eq.~(\ref{series}) to data on an
interval of $Q^2$. An exploration of this possibility can be meaningfully
carried out in the low-$Q^2$ region at present only on the
$\tau$-data-based model and its covariances. As in the analogous
Pad\'e study in Sec.~\ref{sec3a}, we find
that a representation one order higher is required to reach the same
accuracy for the fitted version as was reached using the corresponding
moment approach. Fitting the coefficients of the cubic form to the
model data at the points $Q^2=0.10,\, 0.11,\, \cdots ,\, 0.20$~GeV$^2$,
for example, yields estimates for $\hat{a}_\mu^{\rm LO,HVP}[Q^2_{max}]$
accurate to between $0.6$ and $0.9\%$ for $Q^2_{max}$ in the interval
from $0.1$ to $0.2$~GeV$^2$. The accuracy of the fitted version
in this case, though good, is less so than what was achieved for the analogous
$[2,1]_H$ Pad\'e fit.  The Pad\'e approach may thus be favored
if one is forced to fit coefficients using data
over a limited range of $Q^2$, while the conformal approach will
be most useful
if high-accuracy determinations of the time moments, and hence the
derivatives of the polarization at $Q^2=0$, turn out to be achievable.
\\

\subsection{\label{sec3c}Chiral representations of the subtracted
polarization}
In the region of interest, $Q^2\,\ltap\, 0.2$~GeV$^2$,
$Q^2$ is sufficiently small that ChPT should be capable
of providing
an accurate representation of the subtracted polarization.
It has been known for some time that the next-to-leading-order (NLO)
representation~\cite{gl84,gl85,gk95,abt00} is not
adequate for this purpose, its slope with respect to $Q^2$ being much less
than what is seen in either lattice data~\cite{AB07} or the continuum
version of the $I=1$ subtracted polarization discussed above.
The source of the problem is the absence, in the NLO representation,
of NLO low-energy-constant (LEC) contributions
encoding the large contributions associated with the prominent vector
meson peaks in the relevant spectral functions. These contributions
first appear at NNLO.

The NNLO representation of the subtracted $I=1$ polarization function has
the form~\cite{abt00,gk95}\footnote{Note that Eq.~(19)
of Ref.~\cite{abt00} contains a misprint: there should be no factor
$q^2$ in the term proportional to $(L_9^r+L_{10}^r)$.}
\begin{equation}
\label{chiralrep}
\left[\hat{\Pi}^{I=1}(Q^2)\right]_{\rm NNLO}\, =\, {\cal R}(Q^2;\mu )\, +\,
c_9(Q^2;\mu )\, L_9^r(\mu )\, +\, 8\, C_{93}^r(\mu )\, Q^2\ ,
\label{subpolnnlo}\end{equation}
where $\mu$ is the chiral renormalization scale, $C_{93}^r$ is one of the
renormalized dimensionful NNLO LECs defined in Refs.~\cite{bce99}, and
${\cal R}$ and $c_9$, which also depend on $m_\pi$, $m_K$ and $f_\pi$, are
completely known once $Q^2$, $\mu$, $m_\pi$, $m_K$ and $f_\pi$ are specified.
The NLO LEC $L_9^r(\mu )$ is well known from an NNLO analysis
of $\pi$ and $K$ electromagnetic form factors~\cite{bt02} and we
take advantage of this determination in the exploratory fits to the
$\tau$-based model data below.

In the resonance ChPT (RChPT) approach~\cite{rchpt},
which one expects to represent
a reasonable approximation for vector channels, $C_{93}^r$ is generated
by vector meson contributions. The RChPT result,
$C_{93}^r\sim -{\frac{f_V^2}{4m_V^2}}\simeq\, -0.017$GeV$^{-2}$~\cite{abt00},
where $f_V$ and $m_V$ are the vector meson decay constant and mass, is
expected to be valid at some typical hadronic scale (usually assumed to be
$\mu \sim m_\rho$). This rough estimate is well supported
by the data, and the term proportional to $C_{93}^r$
is, in fact, the dominant contribution to the RHS of
Eq.~(\ref{subpolnnlo}) for $Q^2\sim 0.1$~GeV$^2$.

In the $I=1$ channel, assuming $C_{93}^r$ to be dominated by the $\rho$
contribution, and expanding the $\rho$ propagator to one higher order
in $Q^2$, one obtains an NNNLO contribution of the form $C\, Q^4$ which is
$-Q^2/m_\rho^2$ times the NNLO contribution $8C_{93}^r Q^2$,
yielding $C\, =\, -8C_{93}^r/m_\rho^2 \sim 0.23$~GeV$^{-4}$.
This estimate
leads to a significantly larger curvature of $\hat{\Pi}^{I=1}(Q^2)$ than
predicted by the known lower-order terms and such a larger curvature
is indeed clearly indicated by the low-$Q^2$
behavior of the $\tau$-data-based model for $\hat{\Pi}^{I=1}(Q^2)$.
Contributions to $\hat{\Pi}^{I=1}(Q^2)$ from a $C\, Q^4$ term with
such a value for $C$ already become numerically non-negligible
at $Q^2\sim 0.1$~GeV$^2$.
In order to allow accurate chiral fits over the range of interest,
we thus need to supplement the NNLO representation of Eq.~(\ref{subpolnnlo})
with an additional $C^rQ^4$ term. $C^r$ represents an
effective NNNLO LEC, which is mass-independent at
that order.{\footnote{The mass-independence of $C^r$
would be relevant if one wished to use the results of chiral fits to
physical-mass continuum data to make predictions about the low-$Q^2$
behavior of the subtracted polarization for lattice simulations
corresponding to sufficiently small, but still unphysically heavy,
light-quark masses.}}  We will refer to the NNLO representation augmented
with the $C^rQ^4$ term as the NN$^\prime$LO representation below.

The NN$^\prime$LO representation is governed by
three LECs, $L_9^r$, $C^r_{93}$ and $C^r$, the first of which
is already known to better than 10\%. The relevant question here is whether,
with sufficiently good Euclidean time moments of the vector correlation
function, or low-$Q^2$ data
for its Fourier transform, this form is capable of
producing a representation of $\hat{\Pi}^{I=1}(Q^2)$ accurate
enough to allow a sub-percent evaluation of the contribution to
$\hat{a}_\mu^{LO,HVP}$ from the region $Q^2\,\ltap\, 0.1-0.2$~GeV$^2$.
It turns out that, at present, the low-$Q^2$ errors on data from lattice
simulations are still too large, and the $Q^2$ coverage too sparse, to allow
this question to be sensibly explored using fake data of the type
employed in Ref.~\cite{gmp13}. We thus investigate the systematics
of the NN$^\prime$LO ChPT fit form using the $\tau$-based $I=1$
model following the same approach as employed in Secs.~\ref{sec3a} and
\ref{sec3b} for the Pad\'e approximant and conformal polynomial forms.
In other words, we determine the relevant LECs, and hence the
chiral representation, from the values of the derivatives of
$\hat{\Pi}^{I=1}(Q^2)$ with respect to $Q^2$ at $Q^2=0$ in
the model. As mentioned before, in the lattice context these derivatives can,
in principle, be determined from the time moments of the Euclidean
correlation function.

\begin{figure}[t]
\caption{\label{fig8}Comparison of the results of the NN$^\prime$LO
representation (\ref{chiralrep})
and the $\tau$-data-based model for $\hat{\Pi}^{I=1}(Q^2)$ (solid curve).
The dashed line shows the result including the phenomenological
term $C^rQ^4$, the dotted line the result
with the NNNLO contribution $C^rQ^4$ removed.}
\centering
{\rotatebox{270}{\mbox{
\includegraphics[width=4.25in]
{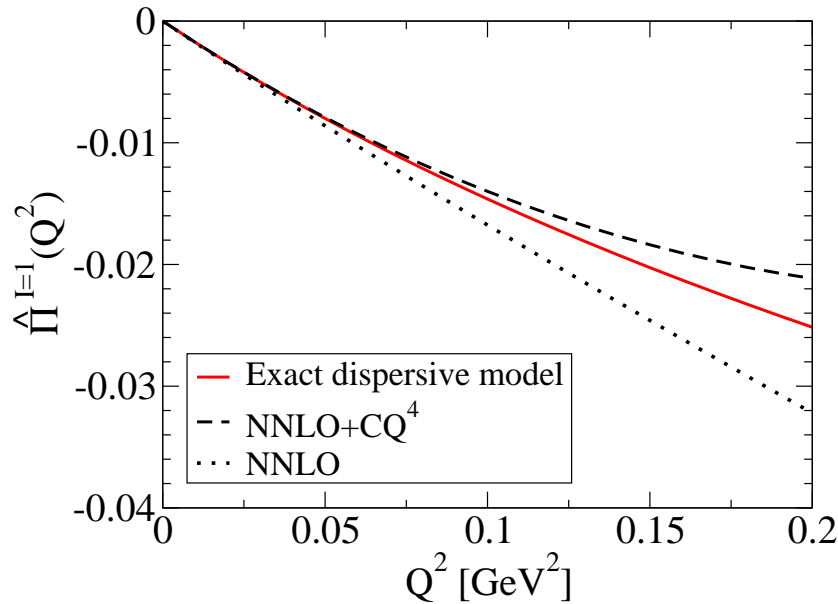}
}}}
\end{figure}

Using $m_\pi=139.57$~MeV, $m_K=495.65$~MeV, $f_\pi=92.21$~MeV,
and $\mu=770$~MeV, as well as $L_9^r(\mu)=0.00593$ from Ref.~\cite{bt02},
and the exact values for ${\Pi^{I=1}}'(0)$ and ${\Pi^{I=1}}''(0)$
from our model,
we find that $C_{93}^r(\mu)=-0.01567$~GeV$^{-2}$ and
$C^r(\mu)=0.2761$~GeV$^{-4}$.\footnote{These are in
rough agreement with the RChPT estimates discussed above. We plan to
present a more detailed discussion of the chiral fits to
$\tau$-decay-based model results for $\hat{\Pi}^{I=1}(Q^2)$ elsewhere.
}
Using these values, Fig.~\ref{fig8} shows the comparison between the exact
model dispersive results for
$\hat{\Pi}^{I=1}(Q^2)$ and those obtained from the chiral representation
(\ref{chiralrep}).
Also shown is the chiral representation
with the $C^rQ^4$ contribution removed. The necessity of the NNNLO
curvature contribution is evident.

Using our chiral representation, we can compare the value for
$\hat{a}_\mu^{\rm LO,HVP}[Q^2_{max}]$ obtained from NN$^\prime$LO
ChPT with the exact-model value.
For $Q^2_{max}=0.1$~GeV$^2$, we find that the ChPT value is $0.6\%$ below
the exact value, while for $Q^2_{max}=0.2$~GeV$^2$, it is $1.4\%$ below.
While the value at $Q^2_{max}=0.1$~GeV$^2$ is acceptable, this is
clearly worse than the approximation obtained using a $[1,1]_H$ Pad\'e
determined from the same derivatives at $Q^2=0$. NNLO ChPT, which
corresponds to setting $C^r=0$, yields values of $4\%$ and $18\%$ above
the exact value, at $Q^2_{max}=0.1$ and $0.2$~GeV$^2$, respectively.
Clearly, the NN$^\prime$LO form provides a good representation for
values of $Q^2$ extending up to about $0.1$~GeV$^2$, but there is
evidence for contributions to the
curvature in the data at higher $Q^2$ beyond that described by the known
NLO, NNLO and $C^rQ^4$ terms. This shows up
in the deviations from the data of the chiral curve
in the region $Q^2\,\gtap\, 0.1$~GeV$^2$ in Fig.~\ref{fig8}.

As in the case of Pad\'e's, an alternative method for constructing a
chiral representation for $\Pi^{I=1}(Q^2)$ is by fits to lattice data at non-zero
values of $Q^2$, instead of from derivatives at $Q^2=0$.   Such fits will
be most reliable when employed in a fit window involving as low $Q^2$ as
possible. {}From Fig.~\ref{fig8} and the discussion above, it follows
that data at values of
$Q^2$ below 0.1~GeV$^2$ would be needed.
In the case of fits to Pad\'e's, we saw in Sec.~\ref{sec3a} that a
sufficiently accurate representation can in principle be obtained from data
in an interval farther away from zero, $0.1\,\ltap\,Q^2\,\ltap\, 0.2$~GeV$^2$ if one
increases the order of the Pad\'e from $[1,1]_H$ to $[2,1]_H$ by adding
one parameter.   In ChPT, such an approach
would imply going beyond NNNLO order.   (As it is, even the NN$^\prime$LO
representation is only a phenomenological version of the NNNLO
representation.)   With such high orders not being available, the
application of ChPT is limited to the moment-based approach, or possibly
to fits at $Q^2$ values below 0.1~GeV$^2$.  This means the ChPT approach to
the low-$Q^2$ region, though potentially providing a consistency check,
is likely to be less useful than the Pad\'e approach. The former
requires small-error data at as low as possible $Q^2$ (something more
difficult to accomplish in practice) while, as shown in Sec.~\ref{sec3a},
a $[2,1]_H$ Pad\'e representation obtained by fitting to
good quality data restricted to the somewhat
higher region of $Q^2$ between approximately $0.1$ and $0.2$~GeV$^2$ can be
employed to obtain a sufficiently accurate value for
$\hat{a}_\mu^{\rm LO,HVP}[Q^2_{max}]$ out to $Q^2_{max}=0.2$~GeV$^2$.
The Pad\'e approach, whether implemented through moments or
through fitting, is thus likely to be a more favorable
one from a practical point of view.

To summarize the conclusions of this subsection, we have shown that,
in the region $0<Q^2\,\ltap\, 0.1$~GeV$^2$, use of NN$^\prime$LO ChPT
provides a representation of the subtracted polarization accurate
enough to allow the evaluation of $a_\mu^{\rm LO,HVP}[0.1$~GeV$^2]$ with
a systematic error at the sub-percent level. Because lattice data at $Q^2$
values below 0.1~GeV$^2$ will be required to reach this level, however,
use of this ChPT-inspired fit form is likely to produce results for
$a_\mu^{\rm LO,HVP}[0.1$~GeV$^2]$ with larger errors than those
obtained from Pad\'e-based approaches.

We conclude this subsection with a brief discussion of the low-$Q^2$
$I=0$ contributions to $a_\mu^{\rm LO,HVP}$. As discussed above, the
NN$^\prime$LO fits to the model $\hat{\Pi}^{I=1}(Q^2)$ data fix
the LECs $C_{93}^r$ and $C^r$. It turns out that at NNLO the related
subtracted vector isoscalar polarization function, $\hat{\Pi}^{I=0}(Q^2)$, is
determined by the same set of LECs as is
$\hat{\Pi}^{I=1}(Q^2)$~\cite{abt00}. This statement
remains true of the NN$^\prime$LO form as well.\footnote{This follows
because contributions of the form $C^r Q^4$ arise
at NNNLO from terms in the effective Lagrangian involving
six derivatives and no quark-mass factors. Such terms will produce
$SU(3)$-flavor-symmetric contributions to the vector current
two-point functions.}
The chiral fit thus also provides us with what should be an accurate
expectation for the behavior of $\hat{\Pi}^{I=0}(Q^2)$ in the
low-$Q^2$ region.  In the isospin limit, $\hat{\Pi}^{I=0}(Q^2)$
determines the $I=0$ contribution to $a_\mu^{\rm LO,HVP}$ via{\footnote{Our
normalization is such that $\hat{\Pi}^{I=0}(Q^2)=\hat{\Pi}^{I=1}(Q^2)$
in the $SU(3)$-flavor limit, with $\hat{\Pi}^{I=1}(Q^2)$ the subtracted
polarization for the flavor $ud$ $I=1$ vector current.}}
\begin{equation}
\left[a_\mu^{\rm LO,HVP}\right]^{I=0}\, =\, -2\alpha^2\,
\int_0^\infty dQ^2\,f(Q^2)\, {\frac{1}{3}}\, {\hat{\Pi}^{I=0}}(Q^2)\ .
\label{amuieq0}\end{equation}

\begin{figure}[t]
\caption{\label{fig9}The NN$^\prime$LO ChPT expectation for
the low-$Q^2$ behavior of the integrand for the $I=0$ contribution
to $a_\mu^{LO,HVP}$. Also shown, for comparison, is the
integrand for the corresponding $I=1$ contribution.}
\centering
{\rotatebox{270}{\mbox{
\includegraphics[width=4.25in]
{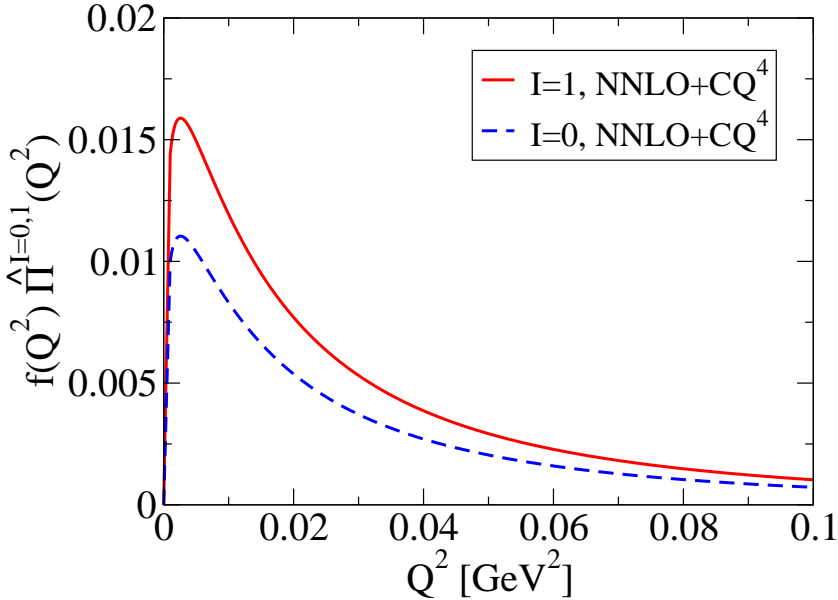}
}}}
\end{figure}

Fig.~\ref{fig9} shows the NN$^\prime$LO expectation for the
product $f(Q^2) \hat{\Pi}^{I=0}(Q^2)$ appearing in the integrand
of Eq.~(\ref{amuieq0}). The corresponding $I=1$ product
$f(Q^2) \hat{\Pi}^{I=1}(Q^2)$ is included for comparison. It is
clear that, though the $Q^2$ dependence of the two is not identical,
the behavior of the $I=0$ integrand is sufficiently similar to that of
the $I=1$ integrand that our conclusions regarding the low-$Q^2$ $I=1$
contribution to $a_\mu^{\rm LO,HVP}$ will also hold for the
$I=0$ contribution.

\section{\label{sec4}Errors for the hybrid strategy and conclusions}

We have shown that the problem of determining the LO HVP contribution to
$a_\mu$ on the lattice can be profitably approached through a hybrid
strategy in which contributions from $Q^2\ge Q_{min}^2$ are evaluated by direct
trapezoid rule numerical integration of lattice data for the subtracted
polarization and those from the low-$Q^2$ region, $0\le Q^2\le Q_{min}^2$, by
other methods. Existing lattice data produced in simulations using
periodic boundary conditions, even without further improvements such
as AMA and/or the use of twisted boundary conditions, are already
sufficiently precise to allow the $Q^2\ge Q_{min}^2$ contributions to be
obtained with systematic and statistical errors well below $1\%$ of
$a_\mu^{\rm LO,HVP}$ for $Q_{min}^2$ as low as $0.1$~GeV$^2$.

In evaluating contributions from the region of $Q^2$ below
$Q_{min}^2\sim 0.1$~GeV$^2$, we have shown, by studying a physical model
of the $I=1$ vector polarization function, that low-order Pad\'e's,
conformally mapped polynomials, as well as
NN$^\prime$LO ChPT (NNLO ChPT supplemented by an additional curvature contribution
whose physical origin is understood) provide forms capable of
representing the subtracted polarization with sufficient accuracy
to reduce the systematic uncertainty arising from computing
$\hat{a}_\mu^{\rm LO,HVP}[Q_{min}^2]$ using these forms to a level well
below $1\%$ of $\hat{a}_\mu^{\rm LO,HVP}$. In the case of the low-order
Pad\'e's, this conclusion remains in force for $Q_{min}^2$ out to
beyond $0.2$~GeV$^2$. In contrast, systematic errors associated with
the use of the NN$^\prime$LO ChPT form grow to about $1.4\%$
of $\hat{a}_\mu^{\rm LO,HVP}$ for $Q_{min}^2\sim 0.2$~GeV$^2$.

A promising approach to the low-$Q^2$ region, from a systematic
point of view, appears to be that involving the Pad\'e's constructed from
the derivatives of the polarization function with respect to $Q^2$
at $Q^2=0$. These derivatives can be obtained from time moments of the
zero-spatial-momentum two-point function~\cite{hpqcd14}. The hybrid
approach allows use of a lower order than would otherwise be
possible, with the $[1,1]_H$ Pad\'e already being sufficient to produce
a systematic error on the determination of $\hat{a}_\mu^{LO,HVP}[Q_{min}^2]$
safely below $1\%$ for $Q_{min}^2$ out to beyond $0.2$~GeV$^2$.
Reducing the order of the Pad\'e employed has the advantage of
reducing the order to which the time moments must be evaluated with
good accuracy, and thus represents a practical advantage in view of the
expectation that light-quark moment errors will grow rapidly with
increasing order.
Constructing the $[1,1]_H$ Pad\'e requires moments only out to
sixth order. In contrast, evaluating the contribution to
$\hat{a}_\mu^{\rm LO,HVP}$ out to $2$~GeV$^2$ with sub-percent accuracy,
would require at least the $[2,2]_H$ Pad\'e, and hence time moments out
to at least tenth order.

We have also shown that a multi-point implementation of the Pad\'e
approach \cite{abgp12}, in which the parameters of the Pad\'e's are fit rather
than obtained from moments, is also feasible. This version
has the advantage that, with sufficiently good data, it can be
successfully implemented using only data from the region of $Q^2$ between
approximately $0.1$ and  $0.2$~GeV$^2$, where lattice data errors are
typically significantly smaller than at lower $Q^2$. To reach
sub-percent accuracy in this implementation, however, requires
going to the $[2,1]_H$ Pad\'e.\footnote{The $[1,1]$ Pad\'e in the notation
of Ref.~\cite{abgp12}.}

The approach using polynomials in the conformally transformed
variable $w$ also looks promising, provided again that
moment evaluations
of the derivatives of $\Pi (Q^2)$ with respect to $Q^2$ at $Q^2=0$
reach a sufficient level of accuracy. If one is forced
to estimate the polynomial coefficients by fitting, however, this
approach looks less favorable than the corresponding Pad\'e approach.

While in principal also usable, the ChPT-based approach appears to us
to require better lattice data to reach the same level of precision
than do the two Pad\'e approaches. This is a consequence of (i) the
necessity of performing the NN$^\prime$LO fits on intervals
restricted to $Q^2\,\ltap\, 0.1$~GeV$^2$ if one wishes to
keep the associated systematic errors  at the sub-percent level, and
(ii) the fact that errors on lattice data are typically significantly
larger below $Q^2\sim 0.1$~GeV$^2$ than they are in the interval
between $0.1$ and $0.2$~GeV$^2$.

Current low-$Q^2$ lattice data are not yet sufficiently precise to produce
sub-percent level statistical errors on the low-$Q^2$ contributions
$a_\mu^{\rm LO,HVP}[Q_{min}^2]$. To understand what might be required to reach
the desired precision, it is convenient to consider the case of
the moment approach, specifically the $[1,1]_H$ Pad\'e representation of
the subtracted polarization,
\begin{equation}
\hat{\Pi}(Q^2)\,=\,\Pi (Q^2)-\Pi (0)\,  =\, {\frac{a_1Q^2}{1+b_1Q^2}}\, ,
\label{pade11form}
\end{equation}
which we know is sufficient to produce systematic uncertainties
well below $1\%$. Errors $\delta a_1$ and $\delta b_1$
on the parameters $a_1$ and $b_1$ produce
associated errors
\begin{eqnarray}
\delta_{a_1}a_\mu^{LO,HVP}[Q_{min}^2]\, &&=\,
-4\alpha^2\, \int_0^{Q_{min}^2}dQ^2\, f(Q^2)\, \left(
{\frac{Q^2}{1+b_1Q^2}}\right)\delta a_1\ ,\nonumber\\
\delta_{b_1}a_\mu^{LO,HVP}[Q_{min}^2]\, &&=\,
-4\alpha^2\, \int_0^{Q_{min}^2}dQ^2\, f(Q^2)\, \left(
{-\frac{a_1Q^4}{(1+b_1Q^2)^2}}\right)\delta b_1\ .
\label{pade11errors}\end{eqnarray}
on $a_\mu^{\rm LO,HVP}[Q_{min}^2]$. Let us now consider the $I=1$ analogue,
for which we can quantify these uncertainties using our $\tau$-data-based
model. Taking the central values for $a_1$ and $b_1$ from
the $[1,1]_H$ Pad\'e version obtained from the derivatives of the
model polarization with respect to $Q^2$ at $Q^2=0$, scaling the
errors, as usual, by $\hat{a}_\mu^{\rm LO,HVP}$, and defining
$c_{a_1}[a_1,b_1,Q^2_{min}]$ and $c_{b_1}[a_1,b_1,Q^2_{min}]$ by
\begin{eqnarray}
{\frac{\delta_{a_1}\hat{a}_\mu^{\rm LO,HVP}[Q_{min}^2]}{\hat{a}_\mu^{\rm LO,HVP}}}
\, &&=\, c_{a_1}[a_1,b_1,Q^2_{min}]\, {\frac{\delta a_1}{a_1}}\nonumber\\
{\frac{\delta_{b_1}\hat{a}_\mu^{\rm LO,HVP}[Q_{min}^2]}{\hat{a}_\mu^{\rm LO,HVP}}}
\, &&=\, c_{b_1}[a_1,b_1,Q^2_{min}]\, {\frac{\delta b_1}{b_1}}\ ,
\label{scalederrordefns}\end{eqnarray}
we find, for example, that
\begin{eqnarray}
\label{cs}
c_{a_1}[a_1,b_1,\, 0.1\ \mbox{GeV}^2] &=&0.818\ ,\nonumber\\
c_{b_1}[a_1,b_1,\, 0.1\ \mbox{GeV}^2] &=&-0.0488\ ,
\end{eqnarray}
and
\begin{eqnarray}
c_{a_1}[a_1,b_1,\, 0.2\ \mbox{GeV}^2] &=&0.913\ ,\nonumber\\
c_{b_1}[a_1,b_1,\, 0.2\ \mbox{GeV}^2] &=&-0.0724\ .
\end{eqnarray}
It follows that a sub-percent error on $a_1$ will be sufficient to
obtain a sub-percent error on $\hat{a}_\mu^{\rm LO,HVP}[Q_{min}^2]$
for $Q_{min}^2\le 0.2$~GeV$^2$, provided the errors on $b_1$
remain at the few percent level, regardless of how correlated the fit
parameters $a_1$ and $b_1$ might be.
The parameter $a_1$ is determined by the slope of the
subtracted polarization with respect to $Q^2$ at $Q^2=0$, and
$b_1$ by the ratio of the curvature to the slope.
A useful rule-of-thumb goal emerging from this exercise is thus
that, to reach the sub-percent error level, one should
aim at reducing the error on the slope parameter $a_1$,
whether obtained from the fourth-order time moment, or from fitting,
to the sub-percent level. Further quantitative studies using our $\tau$-based
model will become possible once covariance matrices associated
with AMA-improved data with twisted boundary conditions become
available.   This will allow us to construct fake data sets based on the
model but with realistic errors and correlations from the point of view of the
lattice.

\begin{acknowledgments}
We like to thank Christopher Aubin, Tom Blum and Taku Izubuchi, as
well as other participants of the Mainz Institute of Theoretical Physics Workshop on the muon $g-2$ for useful discussions.   The authors would like to thank the Mainz Institute for Theoretical Physics (MITP) for its hospitality and support, and KM and SP thank the Department of Physics and Astronomy of
San Francisco State University for hospitality as well.
MG is supported in part by the US Department of Energy, KM is supported by
a grant from the Natural Sciences and Engineering Research Council of Canada,
and SP is supported by CICYTFEDER-FPA2011-25948, SGR2009-894, and
the Spanish Consolider-Ingenio 2010 Program CPAN (CSD2007-00042).
\end{acknowledgments}

%%%%%%%%%%%%%%%%%%%%%%%%%%%%%%%%%%%%%%%%%%%%%%%%%%%%%%%%%%%%%%%%%%%%%%%%%

\vfill\eject
\end{document}